\documentclass{ws-ijmpcs}

\def\be{\begin{equation}}
\def\ee{\end{equation}}
\def\bea{\begin{eqnarray}}
\def\eea{\end{eqnarray}}
\def\del{\partial}

\def\a{\alpha}
\def\b{\beta}
\def\g{\gamma}

\def\k{\kappa}
\def\p{\psi}
\def\P{\Psi}

\def\d{\delta}
\def\D{\Delta}
\def\t{\theta}

\def\l{\lambda}
\def\L{\Lambda}

\def\hA{\hat{A}}
\def\hD{\hat{D}}
\def\hF{\hat{F}}
\def\hP{\hat{\Psi}}

\def\hL{\hat{\Lambda}}

\def\hd{\hat{\delta}}
\def\hs{\hat{s}}
\def\hC{\hat{C}}

\begin{document}

\markboth{Kayhan \"Ulker}
{On the All Order Solutions of Seiberg-Witten Map}

%
\catchline{}{}{}{}{}
%

\title{On the All Order Solutions of Seiberg-Witten Map \\ for Noncommutative Gauge Theories}

\author{KAYHAN \"ULKER}

\address{Abbasa\u ga Mah., Be\c sikta\c s, 
Istanbul, 34353 Turkey\\
kayhan.ulker@gmail.com}

\maketitle

\begin{history}
\received{Day Month Year}
\revised{Day Month Year}
\end{history}

\begin{abstract}

We review the recursive solutions of the Seiberg--Witten map to all orders in $\theta$  for gauge, matter and ghost fields. We also present the general structure of the homogeneous solutions of the defining equations. Moreover, we show that the contribution of the first order homogeneous solution to the second order can be written recursively similar to inhomogeneous solutions.

\end{abstract}

\ccode{PACS numbers: 11.10.Nx, 11.15.-q}


\section{Introduction}
   
Noncommutative (NC) gauge theories have found many applications ranging from solid state physics to particle physics and also to quantum gravity since the noncommutativity in space--time induces naturally a quantum structure. 

One of the main tool to study these theories is to use the so called Seiberg-Witten (SW) map which is a relation between the fields that are defined on the NC space to their ordinary counterparts defined on the usual commutative space \cite{sw}. Although the existence of such a map is originally derived by taking two different low energy limits of string theory, soon after it was found out by Wess and his collaborators that the map exists between NC and commutative gauge theories in a more general setting without referring to the string theory \cite{mssw}\cdash\cite{jmssw}. This construction is purely algebraic and works for arbitrary gauge groups such as $SU(N)$.

The SW map is a gauge equivalence relation between NC and commutative gauge theories. It can be solved by studying the map at each order of the deformation (noncommutativity) parameter $\t$. As a result, the commutative counterpart of the NC theory contains new interaction and higher derivative terms at each order in $\t$.   
 
At first sight, it can be thought that to study the first order solution of the SW--map is sufficient to obtain the leading contribution, for instance to the NC Standard Model\cite{cjsww}. However, higher order terms are also needed to study the consistency of the NC theory itself, such as renormalizability. Moreover, when NC gravity is considered, the first order contributions  in $\t$ from the SW--map vanish identically. Therefore, one has to know at least the second order solutions. Second order solutions are studied by several authors\cite{fidanza}\cdash\cite{tw}.  These maps are different from each other up to a homogeneous solution with different coefficients\footnote{The second order solution for the gauge field given by Ref.\refcite{moller} has also a typo.} due to the freedom in the solutions \cite{ak}. 
   
On the other hand, the solution of SW--map to {\it{all orders}} in $\t$ for the gauge field, matter field (both in adjoint and fundamental representation) and the gauge parameter is given in Ref.\refcite{bk}. These solutions match with the second order solutions given in Ref.s \refcite{fidanza}--\refcite{tw}. 

The advantage of the solutions given in Ref.\refcite{bk} is that they are given recursively  in terms of lower order solutions and hence it simplifies considerably to study the higher order contributions in $\t$. Recently, it has been shown that these solutions can also be rewritten in a geometric setting and they are compatible with hermiticity and charge conjugation conditions\cite{ac}.   

Our aim in this short note, is first to review the results given in Ref.\refcite{bk} in a slightly different setting\cite{zumetal} and then to comment on the general structure of the homogeneous solutions of the SW--map to all orders. Moreover, we also show that the contribution of the first order homogeneous solution to the second order can be written again in terms of the first order homogeneous solutions, similar to the inhomogeneous case.  


\section{The Seiberg - Witten Map}

The simplest non-commutative space is the deformation of the ordinary Minkowski space with 
a real constant antisymmetric parameter $\t$ :
$$[{x} ^\mu \, ,\, {x} ^\nu ]_* \equiv {x} ^\mu \, *\, {x} ^\nu - {x} ^\nu \, *\, {x} ^\mu = i\t^{\mu\nu} .$$
\noindent The non--commutativity  is realized with the $*$--product of Groenewold--Moyal \cite{moyal} :
\be\label{star}
f(x)*g(x) \equiv exp\left(\frac{i}{2}\t^{\mu\nu} \frac{\del}{\del x^{\mu}} \frac{\del}{\del y^{\nu}} \right) f(x)g(y)|_{y\rightarrow x}
\ee
which is associative. In this space, NC field theories are obtained by replacing the ordinary products with the $*$--product. 

Therefore, the action for NC Yang--Mills (NCYM) on Moyal space is written as
\be\label{hI}
\hat{S}=-\frac{1}{4}Tr \int d^4 x \hat{F}^{\mu\nu} * \hat{F}_{\mu\nu} = -\frac{1}{4}Tr \int d^4 x \hat{F}^{\mu\nu}  \hat{F}_{\mu\nu}
\ee
where $ \hat{F}_{\mu\nu} = \del_\mu \hA_{\nu} - \del_\nu \hA_{\mu} - i[\hA_{\mu} , \hA_{\nu}]_* $
is the NC field strength of the NC gauge field $\hA$. The action (\ref{hI}) is invariant under the NC gauge transformations: 
\be
\hd \hA_{\mu} = \del_{\mu} \hL - i[\hA_{\mu} , \hL]_* \equiv \hD_{\mu}\hL \quad , \quad \hd  \hat{F}_{\mu\nu} = i [\hL , \hat{F}_{\mu\nu}]_* .
\ee
Here, $\hL$ is the NC gauge parameter. One can also  consider a more general action that contains matter\footnote{In this work we will not specify the statistics of $\hP$. Our results hold both for fermionic or bosonic fields.} $\hP$ that transform under the adjoint or the fundamental representation of the gauge group. For instance for the fundamental representation we write
\be
\hd \hP = i \hL *\hP \quad, \quad \hD_{\mu} \hP = \del_\mu \hP - i \hA_\mu * \hP
\ee 

The SW map between NC gauge field $\hA_\mu$ and its ordinary counterpart $A_\mu$ is given as a gauge equivalence relation\cite{sw}
\be\label{swe}
 \hd \hA_\mu (A;\t) = \hA_\mu (A+\d A ;\t) - \hA_\mu (A;\t)  = \d  \hA_{\mu}(A;\t) .
\ee
A similar gauge equvalence relation can be written between NC matter fields $\hP$ and ordinary matter fields $\p$ \cite{jmssw}
\be\label{swep}
\hd \hP(\p , A;\t) =  \d \hP(\p ,A;\t) .
\ee

These equivalence relations dictate the following functional dependence : 
\be\label{funcdep}
\hA_\mu = \hA_\mu (A;\t)\quad ,\quad \hat{\P} = \hat{\P}(\p, A;\t) \quad ,\quad \hL = \hL (\l,A;\t) 
\ee  
where $\l$ is the ordinary (commutative) gauge parameter and $\d $ is the ordinary gauge transformation :
$$
\d A_{\mu} = \del_{\mu} \l - i[A_{\mu} , \l] = D_{\mu}\l  \quad ,\quad \d\p =i\l\p
$$ 

To solve the gauge equivalence relations (\ref{swe}) and (\ref{swep}) order by order in $\t$, one can write $\hA$, $\hP$ and $\hL$ as  power series in $\t$ :
\bea
&\hA_\mu = A_\mu + A_{\mu}^{(1)} + \cdots + A_{\mu} ^{(n)} + \cdots
\quad , \quad \hP = \p +\P^{(1)} + \cdots + \P ^{(n)} +\cdots & \nonumber\\
&\hL = \l +\L^{(1)} + \cdots + \L ^{(n)} +\cdots&
\eea

By solving simultaneously for $\hA_\mu$ and $\hL$, the first order solutions of Eq.(\ref{swe}) for $A^{(1)}_\mu$ and $\L^{(1)}$ are given as  \cite{sw} :
\be\label{a1}
A_{\g}^{(1)} = -\frac{1}{4}\t^{\k\l} \{ A_\k ,\del_\l A_\g + F_{\l\g} \} 
\quad ,\quad
 \L^{(1)} = -\frac{1}{4}\t^{\k\l} \{ A_\k ,\del_{\l}\l \} .
\ee
and for $\P^{(1)}$ of  Eq.(\ref{swep}) as\cite{jmssw} :
\be\label{p1}
\P^{(1)}  = -\frac{1}{4}\t^{\k\l} A_\k  (\del_\l + D_\l)\p 
\ee

However, solving (\ref{swe}) and (\ref{swep}) simultaneously for higher order terms is extremely difficult. To disentangle this difficulty one can generalize the ordinary gauge consistency condition 
$\d_\a \d_\b - \d_\b \d_\a = \d_{-i[\a ,  \b]}$
to the NC case  \cite{jmssw} 
\be\label{equiv}
\d_{\a} \hL_\b - \d_\b \hL_\a =  i[\hL_\a , \hL_\b]_* -
i \hL_{[\a ,  \b]}  .
\ee
Note that  (\ref{equiv}) is an equation only for the gauge parameter $\hL$ and after expanding $\hA$, $\hP$ and $\hL$ in terms of  the noncommutativity parameter $\t$, one can try to find solutions for $A ^{(n)}$, $\P ^{(n)}$ and $\L ^{(n)}$.


\section{The Solution of the Seiberg-Witten Map }

Let us, consider BRST transformations $s$ instead of ordinary gauge transformations  $\d$:
\be
s A_\mu = D_\mu c \quad ,\quad s \p =i c \p \quad,\quad s c = i c *c
\ee
where $c$ is a Grassmanian ghost field carrying ghost number one. 

One can generalize the BRST transformations $s$ to the NC BRST transformations $\hs$ \cite{zumetal}:
\be\label{hs}
\hs \hA_\mu = \hD_\mu \hC \quad ,\quad \hs \hP =i \hC * \hP \quad,\quad \hs \hC = i \hC *\hC
\ee  
where $\hC$ is the NC counterpart of $c$. Since $c$ carries a ghost number one, $\hC$ (and each term in its expansion in $\t$) also carries the same ghost number. As $s$, the NC BRST transformation $\hs$ is nilpotent
\be
\hs^2=0
\ee   
and therefore one can explicitly study the underlying cohomological structure in the NC case\cite{zumetal,max}.  Moreover, by requiring SW--map respects the gauge equivalence (\ref{swe}) 
\be\label{swes}
\hs \hA_\mu (A;\t) = s  \hA_{\mu}(A;\t) .
\quad,\quad
\hs \hP(\p , A;\t) =  s \hP(\p ,A;\t) 
\ee
 it can be shown that the nilpotency of $\hs$ is nothing but the gauge consistency condition (\ref{equiv}) \cite{zumetal}.

To find the SW--maps of $\hC$, $\hA_\mu$ and $\hP$ at each order in $\t$,  one can write from (\ref{hs}) and (\ref{swes})
\bea\label{sn}
s C ^{(n)} &=& i \sum_{p+q+r=n}  C ^{(p)}\, *^r\, C ^{(q)} \\
s A_{\mu} ^{(n)} &=& \del_\mu C ^{(n)}_\a - i \sum_{p+q+r=n}  [A_{\mu} ^{(p)}\, , \, C ^{(q)}_\a]_{*^r} \nonumber\\
s \P ^{(n)} &=& i \sum_{p+q+r=n}  C ^{(p)}\, *^r\, \P ^{(q)} \nonumber
\eea
where superscript $n$ denotes the order of $\t$ in the expansion. Note that for $\t \rightarrow 0$ we recover the ordinary fields $c$,  $A_\mu$ and $\p$  on the commutative space. Here, $*^r$ denotes the r--th term in the expansion of the star product (\ref{star}) :
$$
f(x)*^{r}g(x) \equiv \frac{1}{r!}\left(\frac{i}{2}\right) ^r \t^{\mu_1\nu_1} \cdots \t^{\mu_r\nu_r} \del_{\mu_1}\cdots\del_{\mu_r}f(x) \del_{\nu_1}\cdots\del_{\nu_r}g(x) .
$$

A new operator $\Delta$ can be defined by reorganizing the equations in (\ref{sn}) such that the n-th order term of the respective field only appear on the left hand side of the equalities  \cite{zumetal}:
 \bea\label{Dn}
\D C ^{(n)} \equiv s C ^{(n)} -i \{ c, C ^{(n)}\}  &=&  i \sum_{\genfrac{}{}{0pt}{}{p+q+r=n,}{p,q \not= n }}  C ^{(p)}\, *^r\, C ^{(q)} \label{dcn}\\
\D A_\mu  ^{(n)} \equiv s A_{\mu} ^{(n)} - i[c, A_{\mu} ^{(n)}]&=& \del_\mu C ^{(n)}_\a - i \sum_{\genfrac{}{}{0pt}{}{p+q+r=n,}{p \not= n }}   [A_{\mu} ^{(p)}\, , \, C ^{(q)}_\a]_{*^r} \label{dan}\\
\D \P ^{(n)} \equiv s \P ^{(n)} -ic\,\P ^{(n)} &=& i \sum_{\genfrac{}{}{0pt}{}{p+q+r=n,}{q \not= n }}   C ^{(p)}\, *^r\, \P ^{(q)} \label{dpn}
\eea

The operator $\D$ is also nilpotent 
$$
\D^2 =0
$$
and as it is seen clearly from its definition it carries ghost number. 

Our aim is to solve Eqs.(\ref{dcn},\ref{dan},\ref{dpn}) order by order. As it is shown in Ref.\refcite{bk} these solutions are recursive relations between the lower order solutions and higher order ones.

However, these solutions are not unique. One can always extract the homogeneous part from Eqs. (\ref{dcn},\ref{dan},\ref{dpn}) 
\be\label{homo}
\Delta \tilde{C} ^{(n)}  = 0\quad,\quad \Delta \tilde{A} ^{(n)} _\mu =0  \quad,\quad \Delta \tilde{\P} ^{(n)} =0.
\ee
and add at each order any homogeneous solution $\tilde{C} ^{(n)}  , \tilde{A} ^{(n)} _\mu , \tilde{\P} ^{(n)} $ to the inhomogeneous solutions $C ^{(n)} , A ^{(n)} _\mu , \P ^{(n)} $ with arbitrary coefficients \cite{jmssw}. (This freedom in the solutions were first studied in Ref.\refcite{ak}.) Moreover, adding a homogeneous solution at lower orders will also contribute to the higher orders. We will denote these contributions  as barred fields $\bar C ^{(n)} , \bar A ^{(n)} , \bar \P ^{(n)}$.

\subsection{Inhomogeneous solutions to all orders}

By studying Eqs.(\ref{dcn} -- \ref{dpn}) order by order, the inhomogeneous solutions to all orders  are obtained in Ref.\refcite{bk} as 
\bea
C ^{(n+1)}  &=&  -\frac{1}{4(n+1)}\t^{\mu\nu} \sum_{p+q+r=n} \{ A_{\mu } ^{(p)}, \del_{\nu } C ^{(q)} _\a \}_{*^{r}} \\
A ^{(n+1)}_\mu &=& -\frac{1}{4(n+1)}\t^{\mu\nu} \sum_{p+q+r=n} \{ A_{\mu } ^{(p)},  \del_\nu A ^{(q)} _\g + F ^{(q)}_{\nu\g} \}_{*^{r}} \label{an}\\
\P ^{(n+1)} &=& -\frac{1}{4(n+1)}\t^{\k\l} \sum_{p+q+r=n} A_\k  ^{(p)} {*^{r}} (\del_\l \P  ^{(q)} + (D_\l \P) ^{(q)})\label{np} 
\eea
where 
$$
(D_\mu \P) ^{(n)} = \del \P ^{(n)} -i \sum_{p+q+r=n} A_\mu  ^{(p)} *^r \P ^{(q)} .
$$

A similar expression can also be written for the field strength\cite{bk}
$$
F_{\g\rho} ^{(n)} = -\frac{1}{4(n+1)}\t^{\k\l}\sum_{{p+q+r=n}} \big( \{A_\k ^{(p)} ,\del_\l F_{\g\rho} ^{(q)} + (D_\l F_{\g\rho} )^{(q)} \} -2 \{ F_{\g\k} ^{(p)} , F_{\rho\l} ^{(q)} \}_{*^r} \big)
$$
where 
$$
(D_\l F_{\g\rho} )^{(n)}  = \del_\l F_{\g\rho} ^{(n)} -i \sum_{{p+q+r=n}} [A^{(p)} _\l , F_{\g\rho} ^{(q)}]_{*^r}.
$$ 

For a matter field that transforms under the adjoint representation of the gauge group a similar solution can be found either by studying the respective equation or by simply dimensionally reducing  the Eq.(\ref{an}) from six dimensions to four where the extra two dimensions commute \cite{su}. The solution is then found to be \cite{bk},
\be\label{nap}
\Phi ^{(n+1)} = -\frac{1}{4(n+1)}\t^{\k\l} \sum_{p+q+r=n} \{ A_\k  ^{(p)}  ,  (\del_\l \Phi ^{tq}  + (D_\l \Phi) ^{(q)}) \}_{*^{r}}   .
\ee
where
$$
(D_\mu \Phi ) ^{(n)} =\del_\mu \Phi ^{(n)} -i \sum_{p+q+r=n} [A_\mu  ^{(p)} , \Phi ^{(q)} ]_{*^r} \,  .
$$

Note also that, the solutions (\ref{np}) and (\ref{nap}) can be used both for bosonic and fermionic fields. 

In Ref.\refcite{bk}, it is also shown that these solutions (\ref{an}--\ref{nap}) can be obtained by directly solving the respective Seiberg--Witten differential equation that can be obtained by varying the deformation parameter infinitesimally $\t \rightarrow \t +\d\t $ for gauge fields \cite{sw}
\be
\d\t^{\mu\nu}\frac{\del\hA_\g}{\del\t^{\mu\nu}} = -\frac{1}{4}\d\t^{\k\l} \{\hA_\k ,\del_\l \hA_\g + \hF_{\l\g}\}_*
\ee 
 and for matter fields \cite{bk}
 \be\label{difeqfp}
\d\t^{\mu\nu}\frac{\del\hP}{\del\t^{\mu\nu}}  = -\frac{1}{4}\d\t^{\k\l}  \hA_\k  * (\del_\l \hP + \hD_\l \hP )
\ee
 We refer to the Ref.\refcite{bk} for details.
   

\subsection{Homogeneous solutions at each order}   

From the definition of the operator $\D$ one sees that it commutes with the covariant derivative $D_\mu$ \cite{zumetal} 
\be
[\D ,D_\mu] = 0 
\ee
and hence we get $\D F_{\mu\nu} = \D (\del_\mu A_\nu - \del_\nu A_\mu -i [A_\mu , A_\nu])=0$. 

Solutions of the homogeneous equations at each order for the gauge field and the matter fields
\be
\D \tilde A^{(n)}_\mu = 0 \quad, \quad \D \tilde \P^{(n)} =0 
\ee
can be written at each order by dimensional analysis (and matching the ghost number of the fields) such that these solutions contain appropriate powers of $\theta^{\mu\nu}, D_\mu$ and $F_{\mu\nu}$ :  
\be
\tilde A ^{(n)}_\g \propto \mathcal{F}^{(n)}_\g (\t,  D ,  F)  \quad , \quad \tilde \P ^{(n)}  \propto \mathcal{P}^{(n)}(\t , D , F) \p.
\ee

For instance, at first order we have\footnote{Note that $D_\mu  \t^{\mu\nu} F_{\nu\g} \propto D_\g  \t^{\mu\nu} F_{\mu\nu}$ via Bianchi identity.} 
\be\label{homsol1}
\tilde A^{(1)}_\g = l_A ^{(1)} \t^{\mu\nu} D_\g   F_{\mu\nu} \quad,\quad \tilde \P^{(1)}= l_\p ^{(1)} \t^{\mu\nu} F_{\mu\nu} \p
\ee
and typical second order solutions have the form
\be
\tilde A^{(2)}_\g \propto
\t^{\mu\nu} \t^{\k\l}   D_\g ( F_{\mu\nu}  F_{\k\l}) \, , \, \t^{\mu\nu} \t^{\k\l}  D_\k ( F_{\mu\nu}  F_{\g\l} ) \, , \, \cdots
\ee
\be
\tilde \P^{(2)}_\g \propto \t^{\mu\nu} \t^{\k\l}   ( F_{\mu\nu}  F_{\k\l}) \p \, , \,   i \t^{\mu\nu} \t^{\k\l}   (D_\mu F_{\k\nu})D_\l \p \, , \, \cdots
\ee

Since these solutions contain only covariant derivatives and field strengths it is trivial to show that they satisfy (\ref{homo}).

 
\subsection{Contribution of the first order homogeneous solutions to the second order}    

It is clear from the equations that define the all order solutions (\ref{dan},\ref{dpn}), the homogeneous solutions of lower orders will contribute to the higher order ones. In order to find these contributions for the second order let us decompose the fields at the first order fields as $A^{(1)} \rightarrow A^{(1)} + \tilde A^{(1)}$ and $\P^{(1)} \rightarrow \P^{(1)} + \tilde \P^{(1)}$ and the second order ones as $A^{(2)} \rightarrow A^{(2)} + \bar A^{(2)} + \tilde A^{(2)}$ and $\P^{(2)} \rightarrow \P^{(2)} + \bar \P^{(2)} + \tilde \P^{(2)}$. Here, $A^{(1,2)}$ and $\P^{(1,2)}$ are the inhomogeneous solutions,  $\tilde A ^{(1,2)}$ and $\tilde \P ^{(1,2)}$ are the homogeneous solutions and $\bar A^{(2)}$ and $\bar \P^{(2)}$ are the contribution of the first order homogeneous solutions to the second order. 

We can then obtain Eqs. for $\bar A^{(2)}$ and $ \bar \P^{(2)}$ from  Eqs. (\ref{dan},\ref{dpn}) as
\bea
\D \bar A_\g ^{(2)} = i [C^{(1)} , \tilde A_\g ^{(1)}] -\frac12 \t^{\k\l}\{\del_\k c , \del_\l \tilde A_\g^{(1)} \} \\
\D \bar \P^{(2)} = i C^{(1)} \cdot \tilde \P^{(1)} - \frac12 \t^{\k\l} \del_\k c \cdot \del_\l \tilde \P^{(1)} . 
\eea

One can show that $\D^2=0$ still holds by using for instance $\D \del_\l \tilde A_\g^{(1)} = i[\del_\l c, \tilde A_\g^{(1)} ] $ and 
$\D C^{(1)} =  - \frac12 \t^{\k\l} \del_\k c \del_\l c $. Moreover, $\bar A_\g ^{(2)}$ and $\bar \P^{(2)}$ can be obtained in terms of first order homogeneous solutions (\ref{homsol1}) :
\bea
\bar A_\g ^{(2)} = -\frac14 \t^{\k\l} (2 \{A_\k, \del_\l \tilde A_\g^{(1)} \} - i\{A_\k, [A_\l ,  \tilde A_\g^{(1)}] \}  )\\
\bar \P ^{(2)} = -\frac14 \t^{\k\l} A_k (2\del_\l  \tilde \P^{(1)} -i A_\l \cdot  \tilde \P^{(1)} ). 
\eea

We hope to present the general structure and the physical implications of the general solutions including the ghost fields $ \tilde C^{(n)}, \bar C^{(n)}$ in a future publication.

\begin{center}
{\bf {Acknowledgments:}}
\end{center}
As (hopefully not) the last director of Feza Gursey Institute, which is still effectively closed now, I dedicate this work to the memory of Prof. Y. Nutku, the first director of Feza Gursey Institute. 

I also thank to the organizers for the nice and stimulating atmosphere at the Julius Wess 2011 workshop where I had the chance to elaborate on some of our old results. This work is supported by Turkish Academy of Science under Young Scientist Program (TUBA-GEBIP).



\begin{thebibliography}{99}
\addcontentsline{toc}{section}{References}


\bibitem{sw}
N.Seiberg and E.Witten, \emph{``String Theory and Noncommutative
Geometry"}, JHEP {\bf 09}, 032 (1999), arXiv:hep-th/9908142.

\bibitem{mssw}
J.~Madore, S.~Schraml, P.~Schupp and J.~Wess,
  ``Gauge theory on noncommutative spaces,''
  Eur.\ Phys.\ J.\  C {\bf 16}, 161 (2000), 
  arXiv:hep-th/0001203.


\bibitem{jssw}
B.~Jurco, S.~Schraml, P.~Schupp and J.~Wess,
  ``Enveloping algebra valued gauge transformations for non-Abelian gauge
  groups on non-commutative spaces,''
  Eur.\ Phys.\ J.\  C {\bf 17}, 521 (2000), 
  arXiv:hep-th/0006246.

\bibitem{jmssw} 
B.~Jurco, L.~Moller, S.~Schraml, P.~Schupp and J.~Wess,
  ``Construction of non-Abelian gauge theories on noncommutative spaces,''
  Eur.\ Phys.\ J.\  C {\bf 21}, 383 (2001), 
  arXiv:hep-th/0104153.
  
 \bibitem{cjsww}
X.~Calmet, B.~Jurco, P.~Schupp, J.~Wess and M.~Wohlgenannt,
  ``The standard model on non-commutative space-time,''
  Eur.\ Phys.\ J.\  C {\bf 23}, 363 (2002), 
  arXiv:hep-ph/0111115.


\bibitem{fidanza}
  S.~Fidanza,
  ``Towards an explicit expression of the Seiberg-Witten map at all orders,''
  JHEP {\bf 0206}, 016 (2002), 
  arXiv:hep-th/0112027.
  
\bibitem{moller} 
L.~Moller,
  ``Second order of the expansions of action functionals of the  noncommutative
  standard model,''
  JHEP {\bf 0410}, 063 (2004)
  , arXiv:hep-th/0409085.

\bibitem{hakikat} 
M.~M.~Ettefaghi and M.~Haghighat,
  ``Lorentz Conserving Noncommutative Standard Model,''
  Phys.\ Rev.\  D {\bf 75}, 125002 (2007), 
  arXiv:hep-ph/0703313.

\bibitem{ana}
 A.~Alboteanu, T.~Ohl and R.~Ruckl,
  ``The Noncommutative Standard Model at O(theta**2),''
  Phys.\ Rev.\ D {\bf 76}, 105018 (2007), 
  arXiv:0707.3595 [hep-ph].

\bibitem{tw}
 J.~Trampetic and M.~Wohlgenannt,
  ``Comment on the 2nd order Seiberg-Witten maps,''
  Phys. Rev. D {\bf 76}, 127703 (2007),
   arXiv:0710.2182 [hep-th].
  
\bibitem{ak}
T.~Asakawa and I.~Kishimoto,
  ``Comments on gauge equivalence in noncommutative geometry,''
  JHEP {\bf 9911}, 024 (1999),
   arXiv:hep-th/9909139.
  
  \bibitem{bk}
  K. Ulker and B. Yapiskan,  ``Seiberg-Witten maps to all orders,"
  Phys. Rev. D {\bf 77}, 065006 (2008),
   arXiv:0712.0506v1 [hep-th]
  
  \bibitem{ac}
  P. Aschieri and L. Castellani, ``Noncommutative gravity coupled to fermions: second order expansion via Seiberg-Witten map," 	  arXiv:1111.4822v1 [hep-th].
  
  \bibitem{zumetal}
D. M. Brace, B. L. Cerchiai, A. F. Pasqua, U. Varadarajan, and B. Zumino, ``A cohomological approach to the non-abelian Seiberg-Witten map", JHEP {\bf 06}, 047 (2001), 
arXiv:hep-th/0105192 ;
B. L. Cerchiai, A. F. Pasqua and B. Zumino, ``The Seiberg-Witten Map for Noncommutative Gauge Theories,",
 arXiv:hep-th/0206231v2.
 
  \bibitem{moyal}
J.E. Moyal, \emph{``Quantum mechanics as statistical theory"}, Proc. Camb. Phil. Soc. {\bf 45},  (1949)  99.

\bibitem{max}
  G.~Barnich, M.~A.~Grigoriev and M.~Henneaux,
 ``Seiberg-Witten maps from the point of view of consistent deformations of gauge theories,''
  JHEP {\bf 0110}, 004 (2001),
   arXiv:hep-th/0106188;
 G.~Barnich, F.~Brandt and M.~Grigoriev,
  ``Seiberg-Witten maps and noncommutative Yang-Mills theories for arbitrary gauge groups,''
  JHEP {\bf 0208}, 023 (2002),
  arXiv:hep-th/0206003; 
  G.~Barnich, F.~Brandt and M.~Grigoriev,
 ``Local BRST cohomology and Seiberg-Witten maps in noncommutative Yang-Mills theory,''
  Nucl.\ Phys.\ B {\bf 677}, 503 (2004),
   arXiv:hep-th/0308092.
  
\bibitem{su}
 E.~Ulas Saka and K.~Ulker,
  ``Dimensional reduction, Seiberg-Witten map and supersymmetry,''
  Phys.\ Rev.\  D {\bf 75}, 085009 (2007)
  , arXiv:hep-th/0701178.


\end{thebibliography}
\end{document}